\documentclass[twocolumn,showpacs,amsmath,amssymb,floatfix,superscriptaddress,aps,pre]{revtex4-1}
\pdfoutput=1
\usepackage{graphicx,color}
\usepackage{bm,bbold,empheq,mathdots}
\usepackage{units,enumitem}
\usepackage{hyperref}

\newcommand{\ba}{\begin{eqnarray}}
\newcommand{\ea}{\end{eqnarray}}
\newcommand{\be}{\begin{equation}}
\newcommand{\ee}{\end{equation}}

  \newcommand{\xref}[1]{(\ref{#1})}
  \newcommand{\e}{{\text e}}
 \renewcommand{\d}{{\text{d}}}
  \newcommand{\D}{{\cal D}} 
 \newcommand{\zz}{{\mathbb Z}}
\newcommand{\tr}{\operatorname{tr}}

\begin{document}

\title{%
Elasticity of cross-linked semiflexible biopolymers under tension%
}

\author{Alice von der Heydt}\email{heydt@theorie.physik.uni-goettingen.de}

\author{Daniel Wilkin}%\email{wilkin@theorie.physik.uni-goettingen.de}

\affiliation{Institute for Theoretical Physics, Georg-August University of G\"ottingen, 
Friedrich-Hund-Platz 1, 37077 G\"ottingen, Germany}

\author{Panayotis Benetatos}
\affiliation{Department of Physics, Kyungpook National University, 
80 Daehak-ro, Buk-gu, Daegu, 702-701, Korea}

\author{Annette Zippelius}
\affiliation{Institute for Theoretical Physics, Georg-August University of G\"ottingen,
Friedrich-Hund-Platz 1, 37077 G\"ottingen, Germany}
\affiliation{Max Planck Institute for Dynamics and Self-Organization, Am Fassberg 17, 
37077 G\"ottingen, Germany}

\date{\today}

\begin{abstract}
Aiming at the mechanical properties of cross-linked biopolymers,
%Modelling a pair of biopolymers as 
we set up and analyze a model of two
weakly bending wormlike %semiflexible
chains subjected to a tensile force, %under tension, 
with regularly spaced inter-chain bonds (cross-links) %at equidistant contour positions 
represented by harmonic springs.
Within this model, we compute the force-extension curve and the differential stiffness exactly
and discuss several limiting cases.
Cross-links effectively stiffen the chain pair
by reducing thermal fluctuations transverse to the force and alignment direction. 
The extra alignment due to cross-links
increases both with growing number and with growing strength of the cross-links,
and is most prominent for small force $f$. 
%For chains with a large bending rigidity, our model predicts
%the linear elastic constant to increase considerably upon insertion of a few cross-links only.
For large $f$, the additional, cross-link-induced extension 
%decays asymptotically as $f^{-2}$ for extensible 
%and as $f^{-1}$ for hard cross-links.
is subdominant except for the case
of linking the chains rigidly and
continuously along their contour. 
In this combined limit, we recover asymptotically the elasticity of a weakly bending wormlike chain without constraints, stiffened by a factor four.
%In the linear regime, we discover a different elastic behavior, 
%similar to that of a weakly bending wormlike chain with doubled persistence length.
The increase in differential stiffness can be as large as $100\%$ for small $f$ or large numbers of cross-links.
\end{abstract}

%Valid PACS appear here
%static elasticity, theory in biol. phys.
%biomolecules, mech. properties
%molecule manipulation, proteins and other biol. molec.
%molecular conformation, macromolecules
%82.35.Pq (biopolymers, polymerization)
%36.20.Hb (chemical bonds, macromolecules and polymers)
%
\maketitle

\section{Introduction}

Many important biopolymers, such as DNA, the cytoskeletal filaments 
(filamentous (F-)actin, microtubules, intermediate filaments), 
%combine %feature relatively high bending rigidities, alignment, and inter-filament cross-links.
as well as collagen in the extracellular matrix are fluctuating
macromolecules with a bending stiffness intermediate between that of a
random coil (Gaussian chain) and a rigid rod.
Polymers whose elastic behavior is dominated by their bending rigidity are known as semiflexible.
Numerous experiments probing their elasticity have become available 
\cite{bustam94science_tech,liu-pollack2002}
with the advances in single-molecule manipulation, particularly for DNA.
Intriguing and qualitatively novel mechanical behavior arises
if semiflexible polymers are pairwise permanently cross-linked. %An example
The elasticity of cross-linked biopolymers 
is widely studied experimentally via force-extension measurements.
In this article, we study analytically the force-extension relation 
%for a system of two cross-linked WLCs
of an irreversibly cross-linked pair of semiflexible polymers within a mesoscopic theoretical model.

Ubiquitous as extracellular mechanical support %in the extracellular matrix 
%abundant in multicellular animals
is the connective-tissue protein collagen, %, of characteristic triple-helix structure,  
whose fibrils achieve their strength via 
covalent intermolecular  
%arrange its molecules in a network by tissue-specific
cross-links between triple-helical molecules \cite{tanzer73,eyre2005}. 
Atomic force microscopy~\cite{Graham2004,vdRijt2006,Yang2007} 
has been used to %probe the elasticity of 
analyze single collagen fibrils, which themselves consist of many microfibrils and hence can
be modeled as anisotropic networks of irreversibly cross-linked semiflexible polymers \cite{ben-zippPRL2007}.
Cell shape and stability is provided by the actin cytoskeleton,
a network of cross-linked F-actin
%different topologies and densities, 
ranging in morphology from a dilute mesh to bundles of parallel filaments
\cite{bausch2010rev}.
%filamentous polymers like F-actin form cross-linked bundles 
The special elastic %mechanical 
properties due to cross-linking, closely related to biological function, 
thus have become a subject of increasing interest and vigorous research activity
\cite{gardel-Science2004}.
Yet, theoretical understanding is incomplete and 
explanations based on semi-microscopic descriptions %modeling %theories on the mesoscopic level
are rare. 

Crucial experimental results such as the strong stretching of double-stranded DNA
\cite{bustam-rev2000} have been successfully explained by 
the theoretical force-extension relation for a weakly bending wormlike chain 
\cite{marko_siggia95}.
The wormlike chain (WLC) %model 
\cite{kratky-porod49,Saito1967,grosb-khok} %mimics
maps the conformations of an
inextensible semiflexible polymer 
%governed by bending rigidity $\kappa$
to one-dimensional paths whose statistical weight penalizes curvature and
is determined by two length scales only:
the total contour length $L$ and the directional correlation or persistence length $L_p$, 
proportional to the bending rigidity $\kappa$.
The weakly bending approximation of a WLC \cite{marko_siggia95} 
simplifies analytical treatment by assuming
%a small angle between 
that the tangent vector at any arc-length position and the end-to-end vector %direction
make a small angle.
%along the entire contour. 
This approximation applies to polymers 
with a large persistence length $L_p$ (compared to $L$) 
or subjected to strong stretching.
%The force-extension curve of this model chain \cite{marko_siggia95}      
%%displays a linear regime for small forces $f$ and
%approaches the maximal end-to-end extension $L$ 
%with a characteristic $f^{-1/2}$ saturation at strong stretching.
%in contrast to the $1/f$ saturation of a freely jointed chain. 
%%Distinctive
%Deviations from this behavior in semiflexible polymers may point to
Inhomogeneities or inter-molecular interactions
%whose modeling has 
have been the subject of 
%%as already indicated by 
several modifications and extensions of the weakly bending WLC: 
%%has proven amenable to several modifications
%%which account for polymer inhomogeneities or weak extensibility:
%%served as a starting point for modeling polymer inhomogeneities of several kinds:
In \cite{ben-ulrich-zipp12}, the force-extension relation of two parallel aligned, 
weakly bending WLCs with a single irreversible cross-link
and the elasticity of an anisotropic network of aligned chains
have been analyzed.
%It could be shown that this model qualitatively captures 
%also the behavior of many stretched WLCs with random, permanent cross-links. 
In the wormlike bundle model, an arbitrary number of regularly arranged parallel filaments is effectively cross-linked by a coarse-grained, continuous interaction \cite{claus_WLCbundle2010}.
The effect of %random disorder introduced as 
spontaneous polymer curvature %on the elastic behavior
has been studied in 
%%accounted for
%\cite{ben_ter10,ben_ter101,ben_ter11}. 
\cite{ben_ter10,ben_ter11}.
Weak extensibility of semiflexible polymers at strong stretching has been addressed with a combination of a WLC and a Gaussian chain, the semiflexible harmonic chain (SHC), in \cite{kierfeldSHC2004}.

In this work,
we consider the elasticity of two identical weakly bending WLCs  
connected by an arbitrary number of cross-links regularly spaced along the polymer contour. 
The cross-links are represented by entropic harmonic springs, 
which allow for a finite extent of the inter-polymer distance at the cross-link sites. 
In the case of infinite spring strength, 
we obtain the limit of hard %strong (delta-function-like) 
cross-links (strong topological constraints). 
By introducing infinitely many cross-links at fixed contour length,
we can also model a continuous cross-linking or an attractive inter-molecular interaction. 
%continuum limit in which the spacing between cross-links approaches zero. 
The ladder structure of our system is reminiscent 
of the base-pair sequence of double-stranded DNA,
but for the reversible hydrogen bonding. 
%will call for modeling reversible instead of irreversible cross-links.
%Although the %rope
%ladder structure of our model is reminiscent of the base-pair sequence of double-stranded DNA, 
%in its present form it applies to irreversible rather than to reversible cross-linking.
This obvious modification of our model to reversible and/or sectional cross-linking
may prove versatile %relevant
for future studies of, e.g., the denaturation of DNA.
%filament pair cross-linked along contour \emph{sections\/} only. 

The paper is organized as follows: 
In Section \ref{sec-model}, we introduce the model and the observable. 
In Section \ref{sec-part-fct}, we present the main steps in calculating the canonical partition function,
from which all equilibrium quantities can be derived. 
Details of this calculation are given in Appendices 
\ref{app:form-uuT} and \ref{app:traceCinvUUT^k}.
In Section \ref{sec-force-ext}, we present the central result which is the force-extension relation. 
After presenting the general result, we particularly focus on the limit of hard cross-links,
of continuous cross-linking,
the linear elasticity for small forces, and the strong stretching limit.
For the limit of continuous cross-linking, 
the general result and a short comparative discussion are given in Appendix \ref{cont-lim}.
We conclude and discuss further extensions of this work in Section \ref{sec-discuss}.

\section{\label{sec-model}Model}
\begin{figure}[h]
\includegraphics[width=\columnwidth]{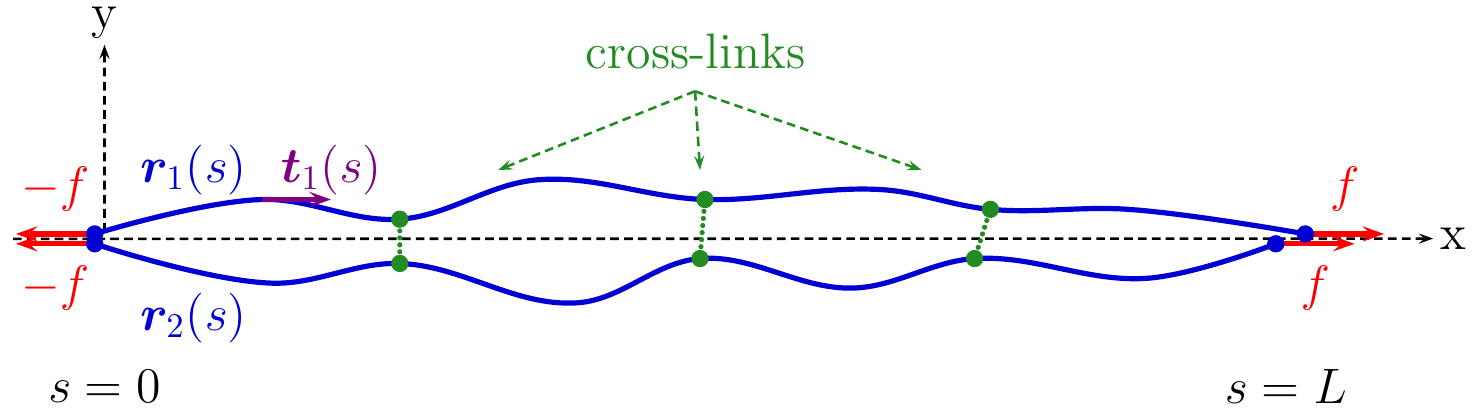}
\caption{\label{model-sketch}(Color online) Stretched, %Sketch of %multiply cross-linked 
weakly bending (see main text for explanation) chain pair connected by $3$ cross-links.}
\end{figure}

%The system we consider comprises
We consider two weakly bending, 
semiflexible chains of equal bending rigidity and contour length,
aligned parallel along a preferential direction $x$ %usually by
%subjected to stretching. 
%%connected at one end 
and 
cross-linked at equidistant arc-length positions specified below. 
The chain configurations are described by paths $\bm r_1(s)$, $\bm r_2(s)$, 
with $s\in[0,L]$ the arc-length parameter,
%contour parameter,
and tangential vectors $\bm t_j(s) = \partial_s \bm r_j \mathrel{\mathop:}= \partial \bm r_j/\partial s$.
 % $\mathrel{\mathop:}= \frac{\partial \bm r_j(s)}{\partial s}$ 
Our setup, taking into account 
%in
space dimension $d=2$ only, is sketched in Fig.~\ref{model-sketch} 
%%%%%%%%%%%
%%%%%%%%%%%
for $3$ cross-links.
As indicated, we assume hinged-hinged boundary conditions, implying
confinement of the vertical positions (here, to $y=0$) and vanishing curvature at the ends.
These boundary conditions are motivated by the following situation:
%In common experiments, 
Experimentally, %a polymer can be stretched by
a tensile force can be applied via 
optical or magnetic tweezers that control the position of beads 
attached to the polymers' ends, 
cf., e.g., \cite{purohit08}.
%Stretching can be accomplished by manipulating beads at the ends of the chains 
Optical tweezers usually restrict %prevent 
the bead's transverse motion, 
but not the rotation,
%In this way, it is possible to exert tangential forces but no moments
so that no moments are exerted at the ends.
Additionally, we assume 
$x_1(0) = x_2(0)$, in order to exclude an overall $x$ shift between the chains.
%(This could be realized e.g.\ by joining initially the chains' ends at both $s=0$ and $s=L$, 
%before applying the symmetric stretching independently to each chain.)
%such that there is no sliding past each other in $x$ direction.

The effective bending potential of semiflexible chains 
(without taking into account torsions \cite{purohit08}) 
is
\be\label{H-bend}
{\mathcal{H}}_{\text{bend}} = \frac{\kappa}{2} 
\int_0^L \! \d s\, \sum_{j=1}^2  \left| \partial_s \bm t_j(s)\right|^2,
%\left| \frac{\partial \bm t_j(s)}{\partial s}\right|^2,
\ee
with the bending rigidity $\kappa$, related to the persistence length $L_p$ 
in $d=2$ via 
\be\label{kappa-L_p} 
\kappa = \frac{1 %d-1
}{2} k_{\text B} T\, L_p, 
\ee
%($d$ the space dimension), 
and with the local inextensibility %of a semiflexible chain 
constraint
\be
|\bm t_j (s)| \equiv 1,\quad s\in[0,L],\quad j=1,2.
\label{inext-cond}
\ee

In order to account for inextensibility in a mathematically tractable way,
we consider the weakly bending approximation: 
The chains' tangents preferentially align with a given direction, here $x$.
A stretching force of strength $f$,
acting on both ends of the chains (cf.\ Fig.~\ref{model-sketch}), 
%both ends of the chain pair are subjected to a longitudinal stretching force $f$, 
is described by the potential
\begin{align}
\label{H-stretch}
{\mathcal{H}}_{\text{stretch}} 
& = 
-f \bm e_{x} \cdot \sum_{j=1}^2 \bigl( \bm r_j(L) - \bm r_j(0) \bigr) 
\\ 
& = 
-f \sum_{j=1}^2 \int_0^L\!\d s\, \partial_s x_{j}(s).
\notag
\end{align} 
For sufficiently large stretching forces or bending rigidities, 
the tilt of the tangent vector away from the $x$ axis is small,
%deflections of the contours in $y$ direction, orthogonal to the force, are small compared 
%to the increments in $x$ direction.
%With this assumption, the parametrization 
%\be
%\bm r_j(s) =\left(
%	      \begin{array}{l} s - \Delta x_j(s) \\ y_j (s) \end{array} 
%	      \right)
%\ee 
%\cite{ben_ter10} and 
so that the condition Eq.~\xref{inext-cond} 
reads approximately
%yield
\be
%1 - \partial_s \Delta x_j (s) 
\partial_s x_{j} (s) = 
1 - \frac{1}{2} \left( \partial_s y_j(s) \right)^2
	      + {\cal{O}} \left( \left( \partial_s y_j(s) \right)^4 \right).
\label{t-approx}\ee
Inserting this expansion into Eqs.~\xref{H-bend} and \xref{H-stretch} 
and discarding all but quadratic terms in derivatives of $y$, we arrive at
the weakly bending approximations of the bending and stretching potentials.
%Recall that weakly bending applies either at strong stretching or at large persistence lengths.

Cross-links between the two chains
are introduced at $N-1$ sites regularly spaced along the contours,
\be\label{sites} 
s_b = \frac{b L}{N},\quad b=1,2,\ldots,N-1,
\ee
%each cross-link between equal arc-length positions on the chains.
dividing the contour length $L$ into $N$ sections, cf.\ Fig.~\ref{model-sketch}.
%Cross-links are introduced as follows: 
%We divide the chains' contour length $L$ into $N$ %equally-sized 
%sections of equal length and connect 
%corresponding sites %junction points 
%on the two chains by $N-1$ cross-links at
%\be\label{sites} 
%s_b = \frac{b L}{N},\quad b=1,2,\ldots,N-1.
%\ee
Explicitly, we model cross-links as entropic,
harmonic springs of strength
$g = 2\, k_{\text B} T / a_c^2$, 
where $a_c^2$ is the temperature-independent squared equilibrium length of one cross-link.
%and effective potential
%\be\label{H-crossl}
%\frac{g}{2} \sum_{b=1}^{N-1} \bigl( y_1(s_b) - y_2(s_b) \bigr)^2.
%\ee 

%\subsection{Energy functional}

%With Eq.~\xref{H-crossl},
Finally, the total effective Hamiltonian
is
\begin{align}
{\cal{H}} = 
&
\underbrace{
\sum_{j=1}^2 \int_0^L \d s 
\left(
\frac{\kappa}{2}  \left( \partial^2_{s} y_j \right)^2
+ \frac{f}{2} \left( \partial_{s} y_j \right)^2
\right)
- 2fL
}_{{\cal H}_0}
\notag\\
& \mbox{} + \frac{g}{2} \sum_{b=1}^{N-1} \bigl( y_1(s_b) - y_2(s_b) \bigr)^2,
\label{H-total}
\end{align}
where ${\cal H}_0$ is the Hamiltonian of the system without cross-linking, 
and the last term $\propto k_{\text B} T$ describes the entropic cross-links 
in weakly bending approximation.
%\footnote{A global $x$ shift between the chains (shearing) could be considered as another degree of freedom and would produce two additional cross-link terms quadratic and linear in the shift. 
%However, this shift does not alter the $x$-extension of the chains, and the variance around zero is $\propto a_c^2/(N-1)$ only.}.

Starting from the concept of harmonic cross-links at discrete sites %contour positions 
$s_b=bL/N$, 
%with given spring constant, 
we will also consider the limit of continuous cross-linking,
achieved by taking
$N\to\infty$ and $\Delta s \mathrel{\mathop:} = L/N \to 0$.
%or introducing an infinite number of regularly spaced cross-links at finite contour length $L$.
In this case, the strength $g$ of a single cross-link has to go to zero, 
such that the total strength $\tilde g \mathrel{\mathop:} = Ng$ remains finite.
%This limit is well-defined %for our system, 
%and actually has the advantage of yielding a closed analytical result for the partition function.
Replacing $\sum_{b=1}^{N-1} \to \frac{N}{L} \int_0^L \! \d s$ 
in the cross-link part of %the Hamiltonian 
Eq.~\xref{H-total}
gives a continuous, harmonic inter-chain attraction of strength $\tilde g/L$,
\be\label{H-cont-X-link1}
{\mathcal H}^{(\text{c})}
 = \frac{\tilde g}{2L} \int_0^L \! \d s\, \bigl( y_1(s) - y_2(s) \bigr)^2.
\ee

%\subsection{Observables}

It is our aim to study the effect of cross-links on the chain elasticity
and hence to
compute the force-extension relation exactly for an arbitrary number of irreversible cross-links.
Thus, the relevant quantity is the average end-to-end extension of one chain in force direction $x$,
\be\label{def-observ} 
\bigl\langle x \bigr\rangle:=
\Bigl\langle x(L) - x(0) \Bigr\rangle_{\cal H} 
= L - \frac{1}{2} \left\langle \int_0^L \d s \left( \partial_s y \right)^2 
\right\rangle_{\cal H},
\ee 
where $\bigl\langle \cdot \bigr\rangle_{\mathcal H}$ denotes the canonical average with
the Hamiltonian of Eq.~\xref{H-total}. 

The force-extension relation of one weakly bending WLC without
cross-links (Hamiltonian ${\mathcal H}_0$), 
first addressed by Marko and Siggia \cite{marko_siggia95}, 
is
\be
\frac{\bigl\langle x %^{(0)} 
\bigr\rangle_{{\mathcal H}_0}}{L}
=
1 - \frac{L}{2 L_p}
\left\{
\frac{\coth \sqrt{f_r}}{\sqrt{f_r}} - \frac{1}{f_r}
\right\},
\label{f-ext-0}
\ee
 in terms of the dimensionless variable
\be\label{def-f_r}
f_r \mathrel{\mathop:}= fL^2/\kappa,
\ee
which is the ratio of stretching energy, $fL$,
and bending energy, $\kappa/L$.
The force-extension curve displays a linear regime for small forces $f$ and
in the limit of strong stretching
approaches the maximal end-to-end extension $L$ 
with a characteristic saturation $\propto f^{-1/2}$.

\section{\label{sec-part-fct}Partition function} %Theoretical analysis

In this section, we detail the calculation of the canonical partition function,
$\mathcal{Z} = \int\!\D[y(s)]\,\e^{-\beta\mathcal{H}[y(s)]}$ (the configurational integral for both chains denoted by $\D[y(s)]$),
which provides access to all equilibrium observables.
%providing all essential %contains all necessary %substantial
%information about our system. 
For the purpose of this work, the end-to-end extension defined in Eq.~\xref{def-observ}
is obtained from $\ln \mathcal{Z}$ or the free energy \footnote{%
There are several possibilities to compute the force-extension
relation of this model %a cross-linked pair of weakly bending chains 
with quadratic-form Hamiltonian. 
In \cite{ben-ulrich-zipp12}, for the case of
a single cross-link, the matrix
%in a suitable basis 
was inverted using the Sherman-Morrison formula, 
an approach which becomes increasingly involved for larger numbers of cross-links.%
}
by differentiation with respect to the force $f$:  
\be\label{observable-Z}
\bigl\langle x \bigr\rangle =
\frac{k_{\text B}T}{2}
\frac{\partial\ln {\mathcal Z}}{\partial f}.
\ee

The first step is to expand the chain configurations $y_j(s)$ in appropriate eigenfunctions.
As mentioned above, we impose hinged-hinged boundary conditions,
which for our system
translate into [$y^{\prime\prime}_j(s) \mathrel{\mathop:}= \partial_s^2 y(s)$]
\begin{align}
\label{hh-bound-cond}
y_j(0) = y_j(L) & = 0 &
y^{\prime\prime}_j(0) 
= y^{\prime\prime}_j(L) & = 0,\quad j=1,2.
\end{align}
According to these boundary conditions, our Fourier-series %normal-mode
ansatz is
\begin{align}
y_1(s) & = \sum_{m=1}^{M} A_m \sin(q_m s), \notag\\
y_2(s) & = \sum_{m=1}^{M} B_m \sin(q_m s),
\end{align}
with wave numbers
\be\label{q-def}
q_m \mathrel{\mathop:}= \frac{m \pi}{L}, \quad m \text{ the mode number},
\ee
and $M$ the largest undulation mode considered within our continuum model 
(roughly, the wave-length resolution is bounded by molecular distances).
With this ansatz, the Hamiltonian ${\mathcal H}$ %Eq.~\xref{H-total} 
can be written as a quadratic form in the coefficient vector 
\be 
\boldsymbol{\Gamma} \mathrel{\mathop:}= (A_1, B_1, A_2, B_2,\ldots)^T.
\ee
Omitting the constant $-2fL$,
\be\label{H-quadr}
\mathcal{H}[\boldsymbol{\Gamma}]
 = \sum_{\ell,\ell'=1}^{M} \Gamma_{\ell} \Bigl( C_{\ell \ell'} + \left(UU^T\right)_{\ell \ell'} \Bigr) \Gamma_{\ell'}, 
\ee
where $C$, due to ${\mathcal H}_0$ of the uncross-linked system, 
is a diagonal matrix ($\otimes$ denotes the Kronecker product),
\begin{align}
\notag
C &= %\underbrace{
\text{diag}(c_1, c_2, \ldots)
%}_{\displaystyle \tilde C} 
\otimes \left( 
\begin{array}{cc}
1  &  0\\
0 & 1
\end{array}\right), \\ 
c_m & \mathrel{\mathop:}= \frac{L}{4} (f + \kappa q_m^2) q_m^2,
\label{C-def}
\end{align}
and $UU^T$ the matrix due to the cross-link Hamiltonian,
cf.\ the second line of Eq.~\xref{H-total}, or Eq.~\xref{H-cont-X-link1}.

The partition function follows as a generalized Gaussian integral over the mode coefficients normalized by $L$, 
\begin{align}
\label{part-fct}
{\mathcal Z} 
&= \int\!\D[\boldsymbol{\Gamma}]\,\e^{- \beta\mathcal{H}[\boldsymbol{\Gamma}]},
\\
\D[\boldsymbol{\Gamma}] 
&\mathrel{\mathop:}= 
\prod_{m=1}^{M} \d \left(\frac{A_m}{L}\right) \d \left( \frac{B_m}{L} \right).
\notag\end{align}
%presented in detail below. 
%In our context,
Returning to the end-to-end $x$-extension introduced in 
Eqs.~\xref{def-observ} and \xref{observable-Z},
we wish to focus primarily on %are particularly interested in 
the cross-link contribution, i.e., 
\be\label{x-link-contrib-ext}
\bigl\langle \Delta x %_{\text{X-link}} 
\bigr\rangle
\mathrel{\mathop:}=
\Bigl\langle x(L) - x(0)\Bigr\rangle_{\cal H} - \Bigl\langle x(L) - x(0) \Bigr\rangle_{{\cal H}_0},
\ee
since the extension %behavior 
$\bigl\langle x(L) - x(0) \bigr\rangle_{{\cal H}_0}$ 
due to thermal fluctuations of uncross-linked weakly bending WLCs only 
is known \cite{marko_siggia95}.
%In order to split off the contribution 
To that end, we write the partition function as 
${\mathcal Z} = {\mathcal Z}_{\text{rel}} {\mathcal Z}_0$,
where ${\mathcal Z}_0$ is the partition function of the uncross-linked system
and address the relative partition function [cf. Eq.~\xref{H-quadr}], 
\begin{align}
\mathcal{Z}_{\text{rel}} 
&\mathrel{\mathop:}= 
\frac{\int\!\D[\boldsymbol{\Gamma}]\,\e^{- \beta\mathcal{H}[\boldsymbol{\Gamma}]}}{\int\!\D[\boldsymbol{\Gamma}]\,\e^{- \beta\mathcal{H}_0[\boldsymbol{\Gamma}]}}
\label{Z-det}\\ &
= \left(\det( {\mathbb{1}} + C^{-1} UU^T) \right)^{-1/2}.
%\notag\\
%& = \exp\left\{- \frac{1}{2} \tr \ln ( \mathbb{1} + C^{-1} UU^T)  \right\}
%\notag
%\\
%& = \exp\left\{- \frac{1}{2} \tr \sum_{k=1}^{\infty} \frac{(-1)^{k+1}}{k} (C^{-1} UU^T)^k  \right\}
\notag
\end{align}
% the ratio of partition functions of the system with and without cross-links.
%the partition function of the cross-linked system relative to that of the uncross-linked system.
This yields the cross-link-induced extra displacement
\be
\bigl\langle \Delta x %_{\text{X-link}} 
\bigr\rangle
= \frac{k_{\text B}T}{2}
\frac{\partial\ln {\mathcal Z}_{\text{rel}}}{\partial f}.
\label{obs-av-ext} 
\ee

\subsection{Finite number of cross-links}

First, we address a finite number $N-1$ of equidistant harmonic cross-links, 
for which the cross-link Hamiltonian is quadratic, but not diagonal in the modes.
The matrix $UU^T$ is a sum of $N-1$ projectors,
%one for each cross-link:
\begin{align}
\label{UUTKronecker}
UU^T &= \frac{g}{2} \sum_{b=1}^{N-1} \left( \bm u_b \otimes \bm u_b^T \right) \otimes
\left(
\begin{array}{rr}
1  &  -1\\
-1 & 1
\end{array}
\right),
\\
\bm u_b^T & \mathrel{\mathop:}=  
(\sin(q_{1} s_b), \sin(q_{2} s_b), \ldots ).
\notag\end{align}
Using the identity $\det \exp A = \exp \tr A$ to expand the determinant in Eq.~\xref{Z-det}, 
$\mathcal{Z}_{\text{rel}}$ is accessible
via traces of powers of the matrix $C^{-1} UU^T$,
\be\label{Z-rel-tr}
\mathcal{Z}_{\text{rel}} = \exp\left\{- \frac{1}{2} \tr \sum_{k=1}^{\infty} \frac{(-1)^{k+1}}{k} (C^{-1} UU^T)^k  \right\}.
\ee
In the trace of a power $k$ of $C^{-1}UU^T$, 
the $2\times 2$ matrices in the Kronecker products, %of $UU^T$ and $C^{-1}$ 
Eqs.~\xref{C-def} and \xref{UUTKronecker},
merely produce a factor $2^k$, which can be computed separately
before performing the trace operation over the mode indices $m_j$.
Hence, we are left with handling the non-diagonal mode-index structure 
of the projector sum Eq.~\xref{UUTKronecker}.
The corresponding matrix of rank $(N-1)$,
\begin{subequations}\label{P-def}
\be\label{P-def1}
P \mathrel{\mathop:}= \frac{2}{N} \sum_{b=1}^{N-1} \bm u_b \otimes \bm u_b^T, 
\ee
with mode indices $m_1, m_2$ has entries
\begin{align}
P_{m_1 m_2}
&=\frac{2}{N} \sum_{b=1}^{N-1} \sin\left( \frac{b m_1 \pi}{N} \right) \sin \left( \frac{b m_2 \pi}{N} 
\right) 
\notag\\
%& = \frac{1}{N} \sum_{b=1}^{N-1} \left\{ \cos\frac{b (m_1-m_2) \pi}{N} - \cos\frac{b (m_1+m_2) \pi}{N} \right\}
%\notag\\ 
& = \delta_{m_1-m_2, 2\zz N} - \delta_{m_1+m_2, 2\zz N},
\label{P-def2}
\end{align}
\end{subequations}
where $\zz$ denotes the set of integers,
such that $P$ is a sparse matrix of block form: 
In each quadratic block of dimension $2N$, nonzero entries appear 
on the diagonal ($+1$) and on one anti-diagonal ($-1$) only. 
For the explicit form of $P$, see Appendix \ref{app:form-uuT}.
The rows/columns display the occupation structure of the 
%Thereby, the
$N-1$ eigenvectors (labeled by $l$ in the following) in the Fourier basis, 
each with nonzero amplitudes only %at the indices of 
for a subset of modes,
%projector sum $P$ selects pairs of modes 
which at all cross-link sites are pairwise in-phase or phase-shifted by 
$\pi$. %relative to each other
%and which do not have nodes at all cross-link sites.
%The quadratic Hamiltonian for equidistant cross-links sorts all modes into 
%establishes $N-1$ equivalence classes, 
%modes being equivalent if they have the same set of phases at the cross-link sites
%or phases of opposite sign respectively. 
Modes indexed by multiples of $N$ have nodes at all cross-link sites, thus
do not contribute to the cross-link energy, and constitute the kernel of $P$.
%The mode classes are labeled by the phase at, e.g., the first cross-link. 

%Tedious, but standard matrix manipulations that rest on 
Due to this special form of the matrix $C^{-1} UU^T$, 
we are able to derive 
%yield 
a closed expression for the trace of any power of $C^{-1} UU^T$ (cf.\ Appendix~\ref{app:traceCinvUUT^k}),  
%We can compute these traces of any power, 
viz.\
\begin{align}
\label{trCinvUUTk}
\lefteqn{
 \left( \frac{gN}{2} \right)^{-k} 
 \tr \left( C^{-1} UU^T \right)^k
}\\
& =\sum_{l =1}^{N-1} \left(
\sum_{\mu=1}^{\infty} \left( c^{-1}_{2(\mu-1) N + l} + c^{-1}_{2 \mu N - l} \right)
\right)^k.
\notag
\end{align}
Here, due to the fast decay with mode number 
of the inverse coefficients $c_m^{-1}$ from Eq.~\xref{C-def}
-- 
basically inverse elastic constants %moduli
for the undulation modes
--
we have extended the summation over modes to a series.
Combining Eqs.~\xref{Z-rel-tr} and \xref{trCinvUUTk}, 
we find that the relative partition function ${\mathcal Z}_{\text{rel}}$ factorizes 
into $N-1$ different ``eigenvector'' %`eigen-mode-class' 
factors,
or equivalently,
\begin{align}
\label{Z-rel-factor}
\lefteqn{
\ln \mathcal{Z}_{\text{rel}} 
}\\
&=  
- \frac{1}{2} \sum_{l=1}^{N-1}\ln \left\{ 
1 + \frac{gN}{2} 
\smash{\sum_{\mu=1}^{\infty}} \left( c^{-1}_{2(\mu-1) N + l} + c^{-1}_{2\mu N - l} \right)
\vphantom{\sum}
\right\}
\notag\\
&= \mathrel{\mathop:}  \sum_{l=1}^{N-1} \ln Z_{l}.
\notag
\end{align}
By inserting the $c_m^{-1}$ into the series,
we obtain for the factors of the partition function, Eq.~\xref{Z-rel-factor},
\begin{align}
Z_{l} \label{eq-Z_b}
& = \left\{  1 + \frac{gL}{Nf} \Bigl( \psi_{l}(0) - \psi_{l} (\delta_f) \Bigr) \right\}^{-1/2},\\
\psi_{l}(\delta_f) 
& = \frac{\displaystyle\frac{\sinh \delta_f}{\delta_f}}{ \cosh \delta_f - \cos \phi_{l}}.
\notag
\end{align}
Herein, we employ the dimensionless variable
\be\label{def-xf} 
\delta_f \mathrel{\mathop:}= \frac{L\sqrt{f}}{N \sqrt{\kappa}}, 
\ee %$\delta_f = \sqrt{f_r}/N$ 
which is the ratio of two lengths: 
The arc-length spacing $L/N$ between cross-links and the 
%modified directional correlation length along the stretched chain
directional `memory' length $\sqrt{\kappa/f}$ 
of the stretched WLC, or the penetration depth of boundary conditions \cite{ben_ter10}.
The dependence of the partition function on the bending rigidity $\kappa$ is via this ratio only.
Additionally, we define %$\text{sinch}\,x \mathrel{\mathop:}= (\sinh x)/x$ and 
the phases specific to the $N-1$ eigenvectors, 
\be\label{phi_b} 
\phi_{l} \mathrel{\mathop:}= \frac{\pi l}{N}.
\ee
We note that the Gaussian statistical weights and the regular cross-link spacing  
simplify enormously, if not enable at all, analytical calculations.

In fact, Eqs.~\xref{Z-rel-factor} and \xref{eq-Z_b} are a central result of our
paper, yielding the exact free energy of two cross-linked, weakly bending WLCs
as $F = F_0 + \Delta F %\sum_{e=1}^{N-1} F_{e}
$,
where $F_0$ is the free energy of the chain pair without cross-links,
\be 
F_0 = - k_{\text B} T \ln Z_0 = 
%- k_{\text B} T \ln \int\!\D[\boldsymbol{\Gamma}]\,\e^{- \beta\mathcal{H}_0[\boldsymbol{\Gamma}]}
k_{\text B}T \sum_{m=1}^{M} \ln \frac{L^2 c_m }{\pi k_{\text B} T},
\ee
(leading to the force-extension relation Eq.~\xref{f-ext-0}),
and the free-energy %?
increment $\Delta F$ due to cross-links
is the sum of the $F_{l} = - k_{\text B} T \ln Z_{l}$ from Eq.~\xref{eq-Z_b}.
%resp., the series Eq.~\xref{}.
Via $\delta_f$, this free energy 
depends on the dimensionless energy ratio $f_r = N^2 \delta_f^2$ introduced in Eq.~\xref{def-f_r}. 
%i.e., the ratio of stretching energy (work done by the external force $f$) 
%to bending energy $\kappa/L$. 
As already mentioned, use of the weakly bending approximation
requires that either the work done by the external force $f$
or the bending energy is large compared to the thermal
energy,
%either a strong forcing or a large bending stiffness is needed 
in order to restrict the transverse fluctuations to be small.
The dependence on the free parameter $f_r$ will be further discussed 
for the cross-link contribution to the force-extension relation.
In addition, the free energy depends on the ratio of cross-link energy to stretching energy,
\be\label{g_f-def}
\epsilon_g \mathrel{\mathop :}= \frac{gL}{Nf}
\ee
and, of course, %via $N$ 
on the number of cross-links, $N-1$.

\subsection{\label{cont-lim-F}Continuous cross-linking via harmonic inter-chain attraction}

Here, we sketch the derivation for an infinite number of regularly spaced cross-links, $N\to\infty$, at finite total strength $\tilde g \mathrel{\mathop:}= Ng$, and for finite contour length $L$.
With the continuous cross-link Hamiltonian Eq.~\xref{H-cont-X-link1} in normal-mode representation,
\begin{align}
\label{H-cont-X-link}
{\mathcal H}^{(\text{c})}
%& = \frac{\tilde g}{2L} \int_0^L \! \d s\, \bigl( y_1(s) - y_2(s) \bigr)^2 \\
& = \frac{\tilde g}{2L} \sum_{m_1,m_2 = 1}^{M} 
\left(A_{m_1} - B_{m_1}\right)\left(A_{m_2} - B_{m_2} \right)
\\
&\quad\times
%\underbrace{
\int_0^L \! \d s\, \sin(q_{m_1} s) \sin(q_{m_2} s)
%}_{=\frac{L}{2} \delta_{m_1 m_2}}
\notag\\
& = \frac{\tilde g}{4} \sum_{m=1}^{M}  \bigl( A_m - B_m \bigr)^2,
\notag
\end{align}
the total Hamiltonian is diagonal with respect to the mode indices.
Thus, we find for the excess free energy due to the inter-chain attraction
a closed expression, again extending the sum over the modes to a series,
\begin{align}
\Delta F^{(\text{c})} 
&= - k_{\text B} T \ln \mathcal{Z}_{\text{rel}}^{(\text{c})} 
\notag\\
& 
=
\frac{k_{\text B} T}{2} \sum_{m=1}^{\infty} \ln \left( 1 + \frac{\tilde g}{2} c^{-1}_{m} \right),
\label{DeltaFcont}
\end{align}
in agreement with performing 
the limit $N\to\infty$ at finite $\tilde g$
in Eq.~\xref{Z-rel-factor}.

\section{\label{sec-force-ext}Force-extension relation}

Using Eqs.~\xref{Z-rel-factor} and \xref{eq-Z_b},
straightforward yet tedious differentiation with respect to $f$
 %we arrive at
yields the force-extension relation
\begin{align}
\frac{\bigl\langle \Delta x \bigr\rangle}{L}
& = 
%\frac{g_e L^2}{2 N L_p^2 f_r^2}
\frac{k_{\text B} T g}{8 N f^2}
\sum_{l = 1}^{N-1} \frac{n_{l}(\delta_f)}{d_{l}(\delta_f, \epsilon_g)},
\label{x-ext-X-links}
\end{align}
with numerator
\begin{subequations}
\label{parts-x-ext}
\begin{align}
\label{num-x-ext}
n_{l}(\delta_f) = & 
\,2\left( \cosh \delta_f - \cos \phi_{l} \right)\\
&\mbox{} - \left( 1 - \cos \phi_{l} \right) \left[ 
3 \frac{\sinh \delta_f}{\delta_f} - \frac{ 1 - \cos \phi_{l} \cosh \delta_f }{\cosh \delta_f - \cos \phi_{l}}
\right]
\notag
\end{align}
and denominator
\begin{align}
d_{l}(\delta_f, &\, \epsilon_g)  = \notag\\
 & \left( 1 - \cos \phi_{l} \right) \left( \cosh \delta_f - \cos \phi_{l} \right) 
\label{denom-x-ext}\\ 
& \mbox{} + %\frac{2 g_e L}{N L_p f_r} 
\epsilon_g
\left[ 
\cosh \delta_f - \cos \phi_{l} - \left( 1 - \cos \phi_{l} \right) \frac{\sinh \delta_f}{\delta_f} 
\right]
\notag
\end{align}
\end{subequations} 
in terms of the length ratio
$\delta_f$ from Eq.~\xref{def-xf}, the ratio $\epsilon_g$ from Eq.~\xref{g_f-def}, 
and the phases $\phi_{l}$ from Eq.~\xref{phi_b}
\footnote{For a single cross-link, $\cos \phi_1 = 0$, and the expressions in Eqs.~\xref{parts-x-ext} simplify considerably, cf.~Eq.~(9) in \cite{ben-ulrich-zipp12}.}.

\begin{figure}[h!]
\includegraphics[width=8cm]{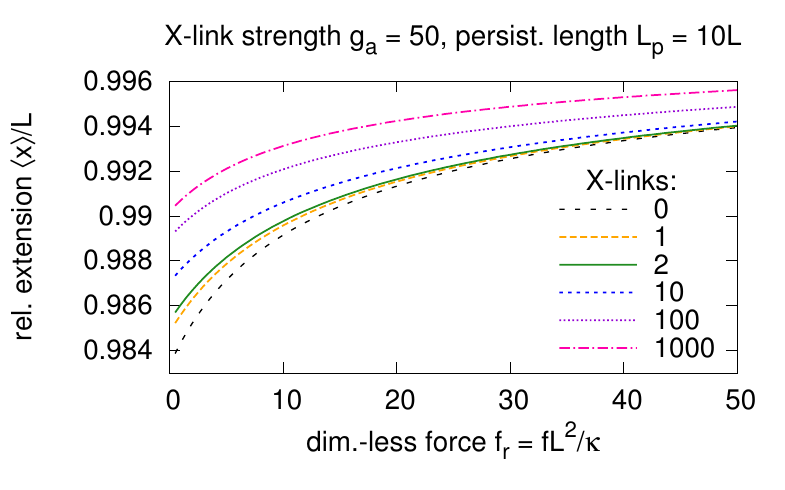}
\includegraphics[width=8cm]{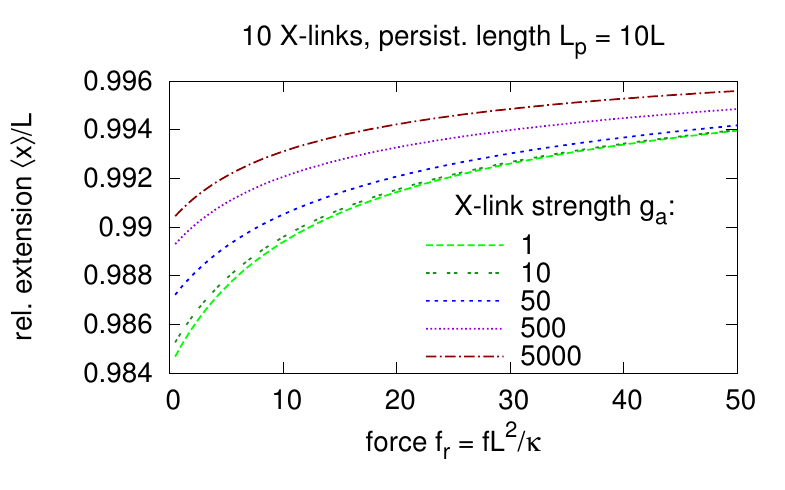}
\caption{\label{fig-Npar_g50}(Color online) Force-extension relation at a relative persistence length 
%to contour length 
$L_p/L=10$: for different numbers of cross-links at finite cross-link strength $g_a = 50$ (top) and for different cross-link strengths at $10$ cross-links (bottom).}
\end{figure}

Since a direct interpretation of the expressions in
Eqs.~\xref{x-ext-X-links} and \xref{parts-x-ext} is difficult,
in Fig.~\ref{fig-Npar_g50}
%%%%%%%%
%%%%%%%%
we show the calculated force-extension relation 
for several numbers and strengths of cross-links, 
as a function of the dimensionless force variable $f_r$, 
%quantifying the ratio of stretching and bending energy, cf.\ 
Eq.~\xref{def-f_r}. 
The dimensionless parameter for the cross-link strength is 
\be 
g_a \mathrel{\mathop:}= \frac{gL^2}{k_{\text B}T} = 2 \frac{L^2}{a_c^2}, 
\ee
which, by virtue of our entropic-spring model for the cross-links,
can be expressed as the squared ratio of the WLC contour length
and one cross-link's length at rest.

First, in the upper part of Fig.~\ref{fig-Npar_g50}, the force-extension curve is plotted
for different numbers of cross-links, at constant strength of a single cross-link.
%Qualitatively, 
The overall form of the saturation curve is %nonlinear in force over a large range,
reminiscent of an unconstrained weakly bending WLC \cite{marko_siggia95}.
Evidently, the general effect of cross-linking is to increase the extension in force direction
relative to an uncross-linked weakly bending chain, 
because cross-links effectively suppress thermal fluctuations perpendicular to the aligning force.
The growth of the extra alignment with the number of cross-links is nonlinear, 
the increase relative to a chain pair with less cross-links being largest for a few cross-links,
and for weak stretching.
%Thus, experimental determination of an increased effective persistence length 
%might
%shed light on the degree of cross-linking, assuming knowledge about the number of cross-linked filaments. 
The limit of continuous cross-linking, cf.\ Sec.~\xref{cont-lim-F}, 
is discussed in Sec.~\ref{subsec-cont-lim}.

Enforcing a smaller and smaller cross-link length (increasing the cross-link strength) 
%results in an increased smoothening of thermal fluctuations, 
enhances the alignment or effective stiffness, too,
as visible in the lower part of Fig.~\ref{fig-Npar_g50}, 
in which the cross-link strength is varied at a fixed number of cross-links.
The limit of strong topological constraints at the cross-link sites (hard or inextensible cross-links)
is presented in Sec.~\ref{subsec-hardXlinks}.

Cross-links are most effective in suppressing transverse fluctuations 
and aligning the chain pair at small reduced stretching forces,
at which the directional memory length $\sqrt{\kappa/f}$ 
is still large compared to the cross-link spacing $L/N$.
For these relatively weak pulling forces,
there is a regime of linear elasticity for all numbers and strengths of cross-links, 
taken a closer look upon in Sec.~\ref{subsec-lin-elast}.
For increasing force, the incremental extension due to cross-links decreases,
since at strong stretching, the dominant contribution to the saturating extension 
arises from ``pulling out'' the remaining length reserves stored in thermal undulations.
The asymptotic decay of the cross-link contribution with force is computed in Sec.~\ref{subsec-ssl}.

In Fig.~\ref{fig-Npar_g50}, 
we have chosen a ratio of persistence to contour length $L_p/L = 10$ 
sufficiently large as to give for all $f_r$ relative extensions close to $1$, 
%consistent with weakly bending approximation
in order to explore the entire range of stretching forces
and yet keep the weakly bending approximation. %for this case
A ratio $L_p/L$ of this order would apply to long microtubules \cite{gittes93microtubules}.
The persistence length of actin is about 15 $\mu$m \cite{gittes93microtubules,ott-LP-actin93}, 
thus for typical lengths of actin filaments in solution, the ratio $L_p/L$ is of order 1.
For smaller ratios, e.g., $L_p/L\sim 0.1$ for type I collagen fibrils \cite{sun2002collagen,bozec2005collagen}, 
or $L_p/L\sim 0.01$ for 10 $\mu$m of double-stranded DNA
\cite{bustam-rev2000}, our predictions are reasonable at strong stretching only.

\subsection{\label{subsec-hardXlinks}Limit 
of hard cross-links}

\begin{figure}[h!]
\includegraphics[width= 7.3cm]{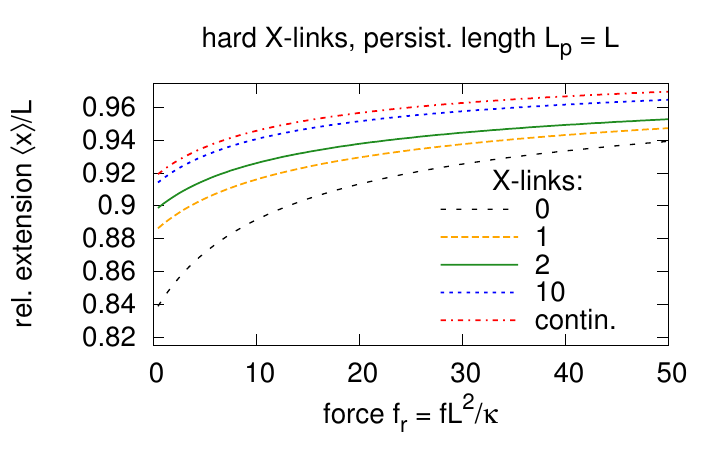}
\caption{\label{fig-hardXlinks}(Color online) Force-extension relation 
for different numbers of hard cross-links %for weakly bending chains with
and $L_p = L$.}
\end{figure}

In the limit of infinite cross-link strength 
or vanishing ratio of cross-link to contour length, 
$a_c/L \to 0$, we have
\be
\frac{\bigl\langle \Delta x^{(\text{h})} %_{\text{hard X-links}} 
\bigr\rangle}{L}
= \frac{k_{\text B} T}{8 f L} %\frac{L}{4L_p f_r}
\sum_{l=1}^{N-1} \frac{n_{l}(\delta_f)}{d^{(\text{h})}_{l}(\delta_f)},
\ee
with $n_l(\delta_f)$ from Eq.~\xref{num-x-ext}, and
\be
d^{(\text{h})}_{l}(\delta_f) =
\cosh \delta_f - \cos \phi_l - \left( 1 - \cos \phi_l \right) \frac{\sinh \delta_f}{\delta_f}.
\ee
The corresponding force-extension curves are shown in Fig.~\ref{fig-hardXlinks},
%%%%%%%%
%%%%%%%%
for more flexible weakly bending chains (with $L_p = L$) than the rather rod-like chains 
in Figs.~\ref{fig-Npar_g50} and \ref{fig-lin-elast}. 
Since hard cross-links completely eliminate relative motion of the filaments transverse to stretching at the cross-linking sites, the relative alignment effect due to cross-links is seen to be stronger.

\subsection{\label{subsec-cont-lim}Continuous cross-linking}

In this Section, 
we discuss the limit of continuous cross-linking from Sec.~\ref{cont-lim-F}, 
$N\to\infty$ 
at total strength $\tilde g \mathrel{\mathop:} = Ng$. 
The incremental extension of one chain due to continuous cross-linking
is computed from $\Delta F^{(\text{c})}$, Eq.~\xref{DeltaFcont}, by differentiation,
cf.\
the general expression for all values of $\tilde g$ and further remarks 
in Appendix \ref{cont-lim}.
In the case of continuous \emph{and\/} rigid cross-linking ($a_c\to 0$), 
the force-extension relation is
\be
\label{cont-rigid}
\frac{\bigl\langle x^{(\text{c, h})} \bigr\rangle}{L}
= 
1 - \frac{L}{4L_p} 
\left\{  
\frac{\coth \sqrt{f_r}}{\sqrt{f_r}}   - \frac{1}{f_r}
\right\}.
\ee
Comparing this result to Eq.~\xref{f-ext-0}, we observe that the 
squared thermal $y$-fluctuations, cf.\ Eq.~\xref{def-observ}, 
are reduced by $1/2$ relative to the uncross-linked case.
This does, however, not imply that the two chains attached to each other rigidly can be treated as unconstrained weakly bending WLCs with just one effective persistence length or bending stiffness $\kappa_{\text{eff}}$, since $f_r$ itself depends on the bending stiffness $\kappa$.
%In the limit of strong stretching, the reduction of thermal fluctuations by a factor two 
%can be interpreted as 
%an apparent stiffness $\kappa_{\text{eff}}= 4\kappa$, 
%whereas for small $f$ to $\kappa_{\text{eff}}=2\kappa$, see below.
The different apparent persistence lengths in the force regimes of linear elasticity and of strong stretching are discussed in the next two Sections.
The extension for rigid, continuous cross-linking for arbitrary force, Eq.~\xref{cont-rigid}, 
is shown as the topmost curve in Fig.~\ref{fig-hardXlinks},
corresponding in Fig.~\ref{fig-Npar_g50} to the asymptotic case 
of both infinite number and strength of cross-links. 
%In this limit, %of rigid, continuous cross-linking, 
%the decay of the extra alignment due to cross-linking with force 
%is slower than for a small number of hard cross-links.
%The expression in Eq.~\xref{cont-lim-ext} reduces
%to the last two terms, which amount to half the difference of the relative $x$-extension and 1
%in the uncross-linked case, cf.\  Eq.~\xref{f-ext-0}, but with opposite sign,
%\be
%\frac{\bigl\langle \Delta x^{(\text{c, h})} \bigr\rangle}{L} = -\frac{1}{2} 
%\left( \frac{\bigl\langle x^{(0)} \bigr\rangle}{L} - 1
%\right).
%\ee
%As a result, the squared $y$-fluctuations, cf.\ Eq.~\xref{def-observ}, 
%are reduced by $1/2$ relative to the uncross-linked case, 
%see also Fig.~\ref{fig-hardXlinks}. 
%Further remarks on the continuum limit and the force-extension relation can be found 
%in Appendix \ref{cont-lim}. 

\subsection{\label{subsec-lin-elast}Force-free extension and linear response regime}

Knowing the exact %analytical expression for the 
extension curve for all values of the force $f$ 
allows us to address the equilibrium extension in the limit $f\to 0$
and the linear elasticity for small $f$ 
-- assuming a large persistence length $L_p$, so that weakly bending holds. 
%to perform the expansion around zero stretching force
%In the force regime, in which the WLC behaves as a Hookean spring, 
This linear response or weak-perturbation regime may be the best accessible
for stretching experiments on sensitive biopolymers.
Moreover, the force-extension curves computed within our model suggest 
that the chain pair's extension at moderate or zero force, %is the best indicator of
cf.\ Fig.~\ref{fig-Npar_g50},
is most indicative of the degree of cross-linking.
Without cross-links, the equilibrium extension of a weakly bending WLC
parallel to alignment for $f\to 0$ is, following Eq.~\xref{f-ext-0},
\be\label{free-ext-0}
\frac{ \bigl\langle x_0 \bigr\rangle_{{\cal H}_0} }{L} 
\mathrel{\mathop:}= \lim_{f \to 0}
\frac{\bigl\langle x \bigr\rangle_{{\cal H}_0}}{L} 
= 1 - \frac{L}{6L_p},
\ee
cf.~\cite{purohit08}.
The deviation from the maximal extension is inversely proportional to the persistence length.
%Even at zero pulling force, a cross-linked WLC displays an
Cross-links increase the equilibrium extension according to
\begin{align}
\label{free-ext-crossl}
\lefteqn{
\frac{\bigl\langle \Delta x_0 \bigr\rangle}{L}
= 
}
\\
& %\frac{k_{\text B} T g L^4}{80 N^5 \kappa^2} 
\frac{g_a L^2}{20 N^5 L_p^2} \sum_{l=1}^{N-1} 
\frac{ \left[ x_l^2 + 13 x_l + 16 \right]/\left( 1 - x_l \right) }{
 6\left(1 - x_l \right)^2 + \frac{2 g_a L}{N^3 L_p} %\frac{gL^3}{N^3 \kappa} 
\left(2 + x_l \right)},
%\notag\\
%& x_l \mathrel{\mathop:}= \cos \phi_{l}
\notag
\end{align}
with $x_l \mathrel{\mathop:}= \cos \phi_{l}$.
%admittedly at first view a little suggestive expression, which apart from the length ratio $L_p/L$ 
%depends on the number $N-1$ of cross-links, the inverse relative cross-link length $L/a_c$, 
%and the eigenvector phases $\phi_l$.
For large $L_p$ and finite $g_a$, this expression is ${\cal O}\left(L/L_p\right)^2$ and hence small.
For hard cross-links, Eq.~\xref{free-ext-crossl} is linear in $L/L_p$,
so that an effective persistence length $L_{p,\text{eff}} > L_p$ 
of the cross-linked chains can be defined, %extracted
viz.,
%\be
%\frac{\bigl\langle \Delta x_0^{(\text{h})} \bigr\rangle}{L}
%= 
%\frac{L}{40 N^2 L_p} \sum_{l=1}^{N-1} 
%\frac{ \left[ x_l^2 + 13 x_l + 16 \right]}{\left( 1 - x_l \right)\left(2 + x_l \right)}
%\ee
\be
\frac{L_p}{L_{p,\text{eff}}} =
1 -
\frac{3}{20 N^2} \sum_{l=1}^{N-1} 
\frac{ x_l^2 + 13 x_l + 16}{\left( 1 - x_l \right)\left(2 + x_l \right)}.
\ee
In the limit of continuous, rigid cross-linking discussed in Sec.~\ref{subsec-cont-lim},
the increase in equilibrium extension is
\be\label{dx0-cont-rigid}
\frac{\bigl\langle \Delta x_0^{(\text{c, h})} \bigr\rangle}{L} = \frac{L}{12 L_p}.
\ee
Upon comparison with Eq.~\xref{free-ext-0}, we thus find
the zero-force extension of one weakly bending WLC with twice the original persistence length
or $\kappa_{\text{eff}}=2\kappa$.

\begin{figure}[h]
\includegraphics[width=7.2cm]{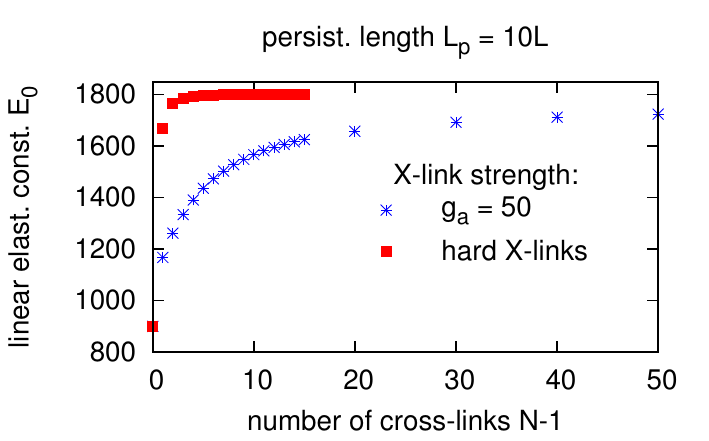}
\caption{\label{fig-lin-elast}(Color online) Linear elastic constant as a function of the number of cross-links, $N-1$, for $L_p/L=10$.}
\end{figure}

%\subsection{Linear response regime}
In Fig.~\ref{fig-lin-elast}, 
%%%%%%%%%
%%%%%%%%%
we show the %highly nonlinear 
dependence of the linear elastic constant,
computed in dimensionless form as
\be
E_0 \mathrel{\mathop:}= \left.\left( \partial_{f_r} \frac{\bigl\langle x \bigr\rangle}{L} \right)^{-1}\right|_{f_r=0}, 
\ee 
on the number of cross-links.
The elastic constant of an uncross-linked weakly bending WLC is given by $E_0 = 90L_p/L$.
For extensible cross-links, a large increase with cross-link number up to $N\approx 10$ 
is followed by a saturation to twice the elastic constant of an uncross-linked weakly bending WLC.
For hard cross-links, the increase in the elastic constant caused by introducing only a few cross-links is even more drastic, and the curve approaches a step function.

%We note that applying a weak tensile force to both chains independently may in experiment 
%induce an additional shearing of the cross-links, not taken into account here,
%which might eventually cause rupture of cross-links. 

\subsection{\label{subsec-ssl}Strong stretching limit}

Here, we consider the limit of strong stretching, i.e., $f_r \gg 1$, or for finite $N$,
$\sqrt{f_r}/N \gg 1$, which means that the directional memory length 
introduced after Eq.~\xref{def-xf} is much smaller than the cross-link spacing,
%yet sufficiently large persistence length as to fulfill
\be
\sqrt{\frac{\kappa}{f}} \ll \frac{L}{N}.
\ee

At finite cross-link strength and for a finite number of cross-links, the dependence on the
individual eigenvector phases remains
in the limit of strong stretching, 
yet the asymptotic scaling with $f_r$ is the same for all summands,
\begin{align}
\lefteqn{
\frac{\bigl\langle \Delta x \bigr\rangle}{L}
}\label{ssl-extens}\\ 
&= \frac{g_a L^2}{N L_p^2} f_r^{-2} \, 
\sum_{l=1}^{N-1} \frac{1}{1 - \cos \phi_{l}} 
+ {\cal O} \left(f_r^{-5/2}\right).
\notag
\end{align}
The same asymptotic decay ensues
for continuous cross-linking at finite total strength $\tilde g_a \mathrel{\mathop:}= 2 NL^2/a_c^2$,
cf.\ Eq.~\xref{cont-lim-ext},
but with a simpler coefficient,
\be
\frac{\bigl\langle \Delta x^{(\text{c})} \bigr\rangle}{L}
= \frac{\tilde g_a L^2}{3L_p^2} f_r^{-2} + {\cal O} \left( f_r^{-5/2}\right).
\ee
%(Note that performing the limit $N\to\infty$ in the coefficient of Eq.~\xref{ssl-extens} erroneously produces a factor 1/2)

In the presence of a finite number of hard cross-links,
strong stretching asymptotically results in an extension increment, 
which is independent of the individual eigenvector phases %of the mode classes 
and decays proportional to $f_r^{-1}$,
\be\label{ssl-hard-Xlinks}
\frac{\bigl\langle \Delta x^{(\text{h})} %_{\text{hard X-links}} 
\bigr\rangle}{L}
= \frac{(N-1) L}{2L_p} f_r^{-1} + {\cal O} \left( f_r^{-3/2} \right).
\ee
%As visible in Fig.~\ref{fig-hardXlinks}
Hence the asymptotic decay of the extra alignment with force is 
markedly slower than for extensible cross-links.
In all cases mentioned so far, the impact of cross-linking diminishes fast for strong stretching,
and the extension curve displays the saturation $\propto f^{-1/2}$
of the uncross-linked chain's extension to the contour length.

Obviously, the asymptotic behavior computed for hard cross-links in Eq.~\xref{ssl-hard-Xlinks} 
cannot apply in the limit $N\to\infty$ of continuous, rigid cross-linking,
due to the diverging prefactor $\propto (N-1)$.
Indeed, in this case, the asymptotic decay of the inter-chain contribution to the extension
is even slower, %cf.\ Sec.~\ref{subsec-cont-lim},
viz.\
\be\label{ssl-cont-lim}
\frac{\bigl\langle \Delta x^{(\text{c, h})} \bigr\rangle}{L}
= \frac{L}{4L_p} f_r^{-1/2} + {\cal O} \left( f_r^{-1}\right).
\ee
%i.e., a decay with the same scaling as the saturation of the uncross-linked chain's extension.
This is the same scaling as the saturation of the uncross-linked chain's extension, hence 
the stabilizing effect of continuous, rigid cross-linking is manifest even for large stretching forces.
Moreover, in the strong stretching limit,
a continuously and rigidly linked chain
behaves effectively like a weakly bending WLC with fourfold original persistence length 
or $\kappa_{\text{eff}}= 4\kappa$.

Of course, the WLC picture is oversimplified at very strong stretching, 
at which inextensibility is clearly violated for many semiflexible biopolymers \cite{smith-overstretchDNA96,liu-pollack2002}.

\subsection{\label{subsec-stiffness}Differential stiffness}

A quantity of interest related to the $x$ extension is the 
differential stiffness: In the corresponding experiment for our system, 
both fibers are pre-stressed by longitudinal stretching, subsequently, 
the strain response to a small change in the applied stress is measured. 
From the force-extension relation, %for all parameter values, 
we can readily compute the differential stiffness 
as the quotient of force %increment needed to produce an 
and extension increment at a given pre-stretching force.
%Mathematically, we obtain this quantity as the inverse slope of the force-extension curve at a value $f_r$.
More precisely, we consider the dimensionless differential stiffness as a function of the dimensionless force $f_r$,
\be
E(f_r) \mathrel{\mathop:}=
\left(\partial_{f_r} \frac{\bigl\langle x\bigr\rangle}{L} \right)^{-1},
\ee
generalizing the elastic constant discussed in Sec.~\ref{subsec-lin-elast}.

\begin{figure}[h!]
\includegraphics[width=7.5cm]{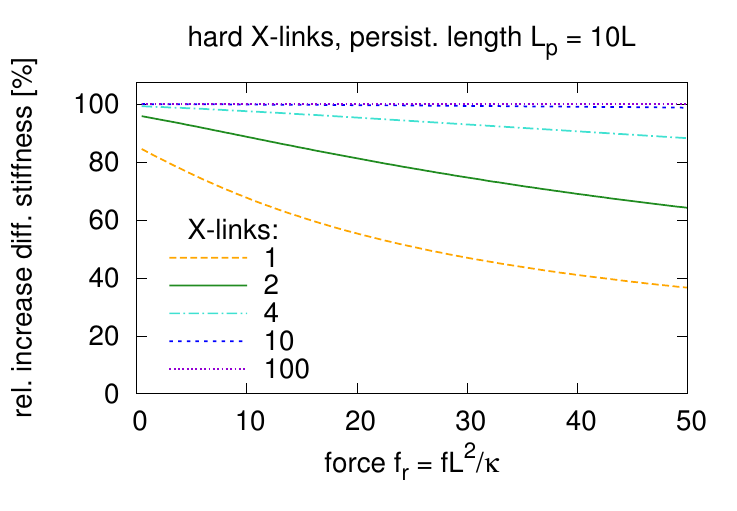}
\caption{\label{fig-stiffn-increase}(Color online) Increase in differential stiffness relative to a weakly bending WLC without cross-links.}
\end{figure}

In an effort to highlight the effective stiffening due to cross-links,
we show in Fig.~\ref{fig-stiffn-increase} 
%%%%%%%%%%%%%
%%%%%%%%%%%%%
the increase in differential stiffness 
relative to uncross-linked weakly bending WLCs for the case of hard cross-links.
Again, for a few cross-links, the differential stiffness is particularly enhanced at weak stretching.
Already for a single, hard cross-link, the linear elastic constant is increased by more than 80 $\%$
compared to the uncross-linked case.
Upon approaching the limit of rigid, continuous cross-linking, cf.\ Eq.~\xref{cont-rigid},
the differential stiffness is increased by a factor two for all values of $f_r$. 
%as mentioned after Eq.~\xref{dx0-cont-rigid}.

\section{\label{sec-discuss}Discussion and Outlook}

%In an effort to study 
Within a transparent mesoscopic model of cross-linked polymers, 
the elasticity of two irreversibly cross-linked
WLCs subjected to a tensile force has been studied in the weakly bending approximation. 
%we have analyzed a stretched WLC pair with equidistant interchain bonds modeled as harmonic entropic springs.
The validity of the latter is granted by assuming either a large tensile force or a large bending rigidity.
For an arbitrary number of cross-links with given strength, 
we have calculated the free energy and, derived thereof, the force-extension relation exactly.
Both with increasing number $N$ and with increasing strength $g$ of the cross-links, 
the effective stiffness of the chain pair increases, 
%or equivalently, transverse fluctuations decrease.
since cross-links stabilize the chains against thermal undulations. 
Particularly for weak stretching, the enhancement in alignment is considerable,
such that in corresponding weak-perturbation experiments on biopolymers,
the increase in the linear elastic constant 
may be a useful indicator of (partial) cross-linking.
As expected, the effect is most pronounced for hard cross-links. 
In the limit of strong pulling forces, the additional extension $\left\langle \Delta x \right\rangle$ 
due to cross-linking decreases, 
and the elasticity of an uncross-linked WLC \cite{marko_siggia95} 
dominates. %as long as the number of cross-links remains finite. 
%impact of cross-links decreases with increasing stretching force parameter.
%We have also considered a situation, in which the chains are linked continuously along their contour.
%The corresponding Hamiltonian allows for the derivation of a closed expression for the partition function.
However, the asymptotic behavior for large stretching forces is different for 
hard and extensible cross-links, as well as for discrete and continuous cross-linking, 
and is summarized in Table~\ref{tab:ssl-exponents}. 
For extensible cross-links, 
the cross-link contribution decays as $f^{-2}$, 
for a finite number of hard cross-links, as $f^{-1}$.
A slower decay is found 
in the limit of both cross-link number $N\to\infty$ and cross-link strength $g\to\infty$,
in which the two chains are linked continuously and rigidly along their contour:  
For strong stretching, the asymptotic form of the force-extension relation reflects the behavior of one uncross-linked weakly bending chain with effective persistence length $4L_p$
and with the known $f^{-1/2}$ scaling. 
%%%%%%%
\begin{table}[h]
\caption{\label{tab:ssl-exponents}Exponents of the asymptotic scaling 
of %extra extension due to cross-links
$\left\langle \Delta x \right\rangle$
with force $f_r = fL^2/\kappa$ 
in the strong stretching limit, $f_r \gg 1$.}
\begin{ruledtabular}
\begin{tabular}{ccc} %{lll} %{@{}*{3}{l}@{}}
%\rule[-2mm]{0mm}{7mm}
%\slashbox{cross-link rigidity}{spacing} 
 & \multicolumn{2}{c}{number of cross-links}	\\
		& finite 	& infinite	\\
%\hline
%\rule[-2mm]{0mm}{7mm}
extensible				& $-2$	& $-2$\\
%\rule[-2mm]{0mm}{7mm}
hard 					& $-1$	& $\nicefrac{-1}{2}$
\end{tabular}
\end{ruledtabular}
\end{table}
%%%%%%%

From the exact extension for all stretching forces, we have computed
another experimentally relevant observable, viz.,  
the differential stiffness of the (pre-stretched) cross-linked chains.
%Again, the impact of cross-linking is largest 
Even a small number of cross-links enhances the differential stiffness dramatically. 
Again, the impact is largest %particularly 
for small stretching forces,
which can be considered within the weakly bending approximation
for WLCs with a large persistence length $L_p/L$.

Several generalizations of our approach are possible: 
%Our approach can be generalized in several ways. For example, 
As alluded to in \cite{ben-ulrich-zipp12}, 
our model is not in principle restricted to a pair of cross-linked filaments, 
but should be generalizable to describe the tensile elasticity of %any two-filament 
a stretched, weakly bending WLC bundle, possibly with random and/or reversible cross-links.
In order to take into account non-affine deformation of cross-linked bundles, 
we may have to consider also the shearing of cross-links.
Detailed %and promising 
analysis of bundles exists, due to the relevance for actin networks, 
mostly for reversibly cross-linked and extensible, semiflexible polymers 
\cite{clausPRL2007,lieleg-clausBundlesPRL2007}.
%In the context of bundled polymer brushes, the effect of grafting the cross-linked filaments to a surface is a further challenging problem to be explored.
Particularly for bundles, the effect of excluded volume interaction, 
neglected in this work, remains to be explored.

Apart from activities in this realm,
%cross-linked semiflexible bundles,
%Recently, a lot of research activity 
major recent research efforts are devoted to the impact of structural inhomogeneities 
caused by the (local) breaking of complimentary base-pair bonds 
(``unzipping'' or denatured ``bubbles'') on the elasticity of double-stranded DNA
\cite{hanke_stretchDNA2008,marenduzzo2009}. 
A class of semi-microscopic models convenient
for analyzing the thermal denaturation transition, as well as the ``bubble'' statistics and dynamics \cite{Theodorakopoulos}, 
focuses on the form of the base-pairing interaction, 
but does not account for the polymers' conformational degrees of freedom,
which determine certain DNA properties \cite{Theo_Peyr2012}. 
In the breathing DNA model \cite{Sung}, 
two discrete chains (consisting of interacting ``beads'') with bending and stretching rigidity
interact via the pairing energy of complimentary bases, represented by a Morse potential. 
Another semi-microscopic model, amenable to a transfer matrix method, 
considers a discrete WLC model for the chain conformations, 
coupled to an one-dimensional Ising model describing the internal base-pair states \cite{Palmeri}.
In the context of denaturation of DNA, 
it would be interesting to extend our model to reversible %instead of irreversible 
cross-linking, in order to study the coexistence of ``ladders'' (cross-linked strand sections) 
and ``bubbles'' (open sections).
 %Cases intermediate between reversible and irreversible cross-linking are imaginable, as well. 
A first, obvious step towards this direction will be to address with our method 
two parallel aligned WLCs
whose arc-length is sectioned into
%with a sequence of 
cross-linked and disconnected parts. %With reversible cross-linking,
In the set-up we have considered here,
all ingredients, viz., cross-linking, bending stiffness, and longitudinal forcing, 
act to decrease transverse fluctuations of the chains.
An unzipping transition could presumably be studied in an altered situation,
%competing forces %that can induce an unzipping transition would be present
e.g., one, in which the cross-linked filaments are teared apart at one end.
More complicated refinement of our model might account for twist and overstretching,
effects shown to be essential for the elasticity of DNA \cite{gross-natphys2011}. 
%For DNA mechanical properties, twisting and overstretching have been shown to be essential , effects that should be incorporated in a refinement of our model.

\begin{acknowledgments}
We thank W.~T.\ Kranz for valuable discussions and careful reading of the manuscript.
Financial support by the Deutsche Forschungsgemeinschaft through grants
SFB-937/A1 and SFB-937/A4 is gratefully acknowledged. 
P.~B.\ acknowledges support by Kyungpook National University Research Fund, 2012.

\end{acknowledgments}

\appendix
\section{\label{app:form-uuT}Structure of the projector sum $P$}
The matrix representation of the cross-link projector sum $P$, cf.\ Eqs.~\xref{P-def}, has the following structure:
\begin{align}
\label{UUT-expl}
P = 
\left(
\begin{array}{ccc}
1 & 1 & \dots\\
1 & 1 & \\
\vdots & &\ddots
\end{array}
\right)
\!\otimes\! 
\underset{2N\times 2N}{
\left(
\begin{array}{cccccccc}
1 & 0 	 & \dots & 0 & \dots & 0 & -1 & 0\\
0 & \ddots &  & \vdots & &\iddots & 0 & \vdots\\
\vdots & 0 & 1 &  & -1 & 0 & \vdots & \\
0 & \dots && 0 &&\dots & 0 &\\
\vdots & 0 & -1 &  & 1 & 0 & \vdots &\\
0 & \iddots &  & \vdots & &\ddots & 0 &\\
-1 & 0 & \dots & 0 & \dots& 0  &1 & \\
0 &\dots & & & & & & 0
\end{array}
\right)
}.
\end{align}

\section{\label{app:traceCinvUUT^k}Trace of powers of %$C^{-1}UU^T$
the projector sum}
By decomposing mode indices $m \in \{1,2,\ldots\}$ according to the block structure into
\begin{align}
\label{index-subs}
\lefteqn{m = 2 \mu N - r \text{ with the definitions } }\\
&% \left\{  
\begin{array}{ll}
\mu 	& \mathrel{\mathop:}= \left\lceil \frac{m}{2N} \right\rceil \in \{1,2,\ldots \} \text{ (block index)},\\ 
r     	& %\mathrel{\mathop:}= 2 \mu N  - m 
\in \{0,\ldots,2N-1\} \text{ (index within a block)},\\ 
\rho (r)	& \mathrel{\mathop:} = \left\lfloor \frac{r}{N} \right\rfloor  \in \{0,1\} \text{ (quadrant within a block)},
\end{array}
%\right.
\notag
\end{align}
the entries of $P$, 
cf.\ Eqs.~\xref{P-def} and \xref{UUT-expl},
can be encoded in product form, 
the first two factors indicating the location of nonzero entries, 
%the zeros at $r_j \in \{0,N\}$, 
the last factor the sign, 
\begin{align} 
P_{m_1 m_2}  & =
\delta_{m_1-m_2, 2 \zz N} - \delta_{m_1+m_2, 2 \zz N} 
\label{P-entry-factor}
\\
& = ( \delta_{r_1, r_2} + \delta_{r_1, 2N-r_2} ) 
%(1 - \delta_{r_1,0}) (1 - \delta_{r_1,N}) 
(1 - \delta_{r_1,\zz N}) (-1)^{\rho_1 + \rho_2}.
\notag
\end{align}
Then, with the diagonal matrix $C$, cf.\ Eq.~\xref{C-def}, 
we write the trace of a power $k$ of $C^{-1}UU^T$, cf.\ Eq.~\xref{UUTKronecker}, as
\begin{align}
\lefteqn{
 \left( \frac{gN}{2} \right)^{-k} 
 \tr \left( C^{-1} UU^T \right)^k
}\notag\\ 
%with summ. convention
%& =
%\underbrace{c^{-1}_{m_1} \left( \delta_{m_1 - m_2, 2 \zz N} - \delta_{m_1 + m_2, 2 \zz N} \right) 
%\cdots 
%c^{-1}_{m_k} \left( \delta_{m_k - m_1, 2 \zz N} - \delta_{ m_k + m_1, 2 \zz N} \right)
%}_{k \text{ factors}} 
%\notag\\
& = \sum_{m_1,\ldots,m_k=1}^{\infty}
c^{-1}_{m_1} \left( \delta_{m_1 - m_2, 2 \zz N} - \delta_{m_1 + m_2, 2 \zz N} \right) 
\cdot \ldots 
\notag\\
& \quad \times c^{-1}_{m_k} \left( \delta_{m_k - m_1, 2 \zz N} - \delta_{ m_k + m_1, 2 \zz N} \right)
\notag\\
& =
\sum_{\mu_1,\ldots,\mu_k=1}^{\infty} %\ldots \sum_{\mu_k=0}^{\infty}
\sum_{r_1,\ldots,r_k = 0}^{2N-1}
%\left( \sum_{r_1 = 1}^{N-1} + \sum_{r_1 = N+1}^{2N-1} \right)\ldots
%\left( \sum_{r_k = 1}^{N-1} + \sum_{r_2 = N+1}^{2N-1} \right)
\label{trCinvUUTkApp}\\
& \quad\times c_{2\mu_1 N - r_1}^{-1} ( \delta_{r_1,r_2} + \delta_{r_1, 2N-r_2} ) 
(1 - \delta_{r_1,\zz N}) %(1-\delta_{r_1,0})(1-\delta_{r_1,N})
(-1)^{\rho_1 + \rho_2} \notag\\ & \quad\times 
c_{2\mu_2 N - r_2}^{-1} ( \delta_{r_2,r_3} + \delta_{r_2, 2N-r_3} ) 
(1 - \delta_{r_2,\zz N})
(-1)^{\rho_2 + \rho_3 } \notag\\
&\quad\times\ldots 
\notag\\
&\quad \times c_{2\mu_k N - r_k}^{-1} ( \delta_{r_k,r_1} + \delta_{r_k, 2N-r_1} ) 
(1-\delta_{r_k,\zz N})
(-1)^{\rho_k + \rho_1}.
\notag
\end{align}
Due to the symmetry of the constraints, % trace operation, 
we are left with the summation over one of the $r_j$ without $0$ and $N$.
If we split this sum %from $0$ to $2N-1$ without $0$ and $N$ 
according to the constraints and to the eigenvector structure mentioned above
as 
\be
\sum_{r = 0}^{2N-1} f_r (1 - \delta_{r,\zz N}) %(1-\delta_{r_1,0})(1-\delta_{r_1,N})
= \sum_{l=1}^{N-1} \left( f_{l} + f_{2N-l} \right),
\ee
%and observe that mixed terms of the form
%sort out expressions such as $\delta_{b_1, N-b_2}$, 
we finally arrive at
\begin{align}
\lefteqn{
 \left( \frac{gN}{2} \right)^{-k} 
 \tr \left( C^{-1} UU^T \right)^k
}\\
& = \sum_{\mu_1,\ldots,\mu_k=1}^{\infty}
\sum_{l=1}^{N-1} 
\left( c_{2(\mu_1-1) N + l}^{-1} + c_{2\mu_1 N - l}^{-1} \right)
\cdot\ldots\notag\\
&\quad \times \left( c_{2(\mu_k-1) N + l}^{-1} + c_{2 \mu_k N - l}^{-1} \right)
\notag\\
& =\sum_{l=1}^{N-1} \left(
\sum_{\mu=1}^{\infty} \left( c^{-1}_{2(\mu-1) N + l} + c^{-1}_{2 \mu N - l} \right)
\right)^k.
\notag
\end{align}
\section{\label{cont-lim}Continuous cross-linking}

Here, we present the general result for the limit of continuous cross-linking 
dealt with in Secs.~\ref{cont-lim-F} and \ref{subsec-cont-lim}.
%Computing the relative partition function with the cross-link Hamiltonian Eq.~\xref{H-cont-X-link} or 
%taking the limit $N\to\infty$ in eq.~\xref{Z-rel-factor} with $\tilde g \mathrel{\mathop:} = Ng$
%yields 
%%%for the relative partition function
%\be\label{z-rel-cont-lim}
%\mathcal{Z}_{\text{rel}}  =
%\prod_{m=1}^{\infty} 
%\left( 1 + \frac{\tilde g}{2} c^{-1}_{m} \right)^{- 1/2}.
%\ee
The incremental extension due to this kind of cross-linking computed from Eq.~\xref{DeltaFcont}
 is %force-extension relation
\begin{align}
\lefteqn{
\frac{\bigl\langle \Delta x \bigr\rangle}{L}
}
\label{cont-lim-ext}
\\
& = \frac{L}{4L_p} 
\left\{  
\frac{\displaystyle%
\frac{\sin g^-}{g^-} - \frac{\sinh g^+}{g^+} }{%
\cosh g^+ - \cos g^-}
%-I*(-x*sinh(y)+y*sin(x))/((cos(x)-cosh(y))*x*y)
%\Biggr.\notag\\&\quad \mbox{} 
+ \frac{\coth \sqrt{f_r}}{\sqrt{f_r}}  - \frac{1}{f_r}
\right\},
\notag
\end{align}
as a function of the dimensionless force $f_r$, Eq.~\xref{def-f_r}, 
and the dimensionless parameters 
\begin{align}
g^{-} & \mathrel{\mathop:}= \sqrt{ 2 \sqrt{2 \tilde g L^3/\kappa} - f_r}, \\
g^{+} &\mathrel{\mathop:}= \sqrt{ 2 \sqrt{2 \tilde g L^3/\kappa} + f_r},
\notag
\end{align}
which apart from $f_r$ contain the ratio of total inter-chain attraction $\tilde g L^2 = 2 N k_{\text B} T L^2/a_c^2$ to bending energy $\kappa/L$.
(The function in
%For any given value of this ratio, %$\tilde g$ and $\kappa$
Eq.~\xref{cont-lim-ext} 
is a continuous, real-valued function of $f_r$ independently of the sign of the outer square root's argument.)

\begin{figure}[h!]
\includegraphics[width=8cm]{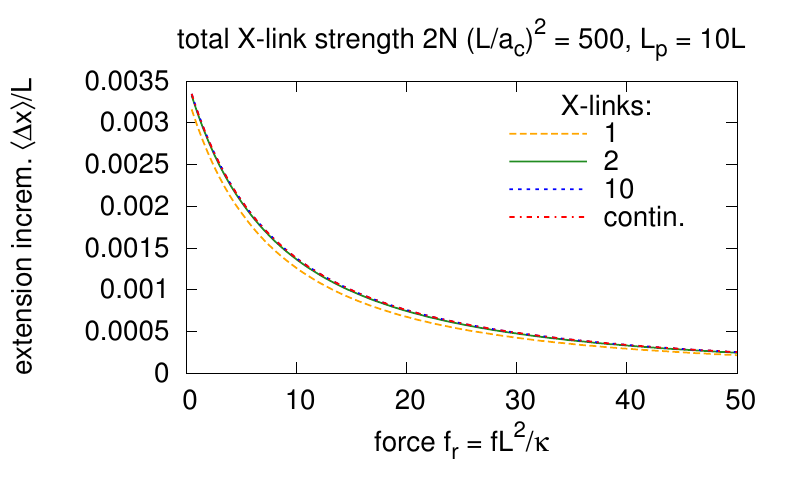}
\caption{\label{fig-cont-lim}(Color online) Cross-link-induced extension increment at total cross-link strength $\tilde g_a=500$ and $L_p/L=10$. %for different numbers of cross-links
}
\end{figure}

Comparing the force-extension curves for finite numbers of cross-links to each other and to those for continuous cross-linking, %at a given persistence length, 
we find an approximate collapse of all curves with the same total 
%cross-linking strength %``amount'' or 
inter-chain attraction $\tilde g_a = 2 NL^2/a_c^2$,
as shown in Fig.~\ref{fig-cont-lim}.
%%%%%%%%%%%%
%%%%%%%%%%%%
Except for the case of a single cross-link, 
the force-extension relation is over a large range of forces to high numerical precision
determined by the product of cross-link number $+1$ and strength of a single cross-link only.
On the basis of Eqs.~\xref{DeltaFcont} and \xref{Z-rel-factor}, %for the free energy, 
this apparent scaling can be traced back 
to the rapid decay of the coefficients $c_m^{-1}$ with mode index $m$. 
%-- inverse moduli for the thermal undulation modes, cf.\ Eqs.~\xref{q-def} and \xref{C-def}. 
The sum over eigenvectors $l$ in Eq.~\xref{Z-rel-factor} for finite $N$ is dominated by those with entries at the lowest modes, 
and of the set of modes represented by one eigenvector, 
only the lowest mode gives an appreciable contribution.
Thereby, the %contribution of the 
``missing''
%multiple-of-$N$ 
modes $m=\zz N$ and, finally, the deviation from the series in Eq.~\xref{DeltaFcont} are negligible 
but for very small $N$.
According to the slower decay $\propto m^{-2}$ of the stretching contribution to the coefficients $c_m^{-1}$, the approximate scaling must break down for strong stretching (and small persistence lengths). In this regime, however, the additional extension due to cross-links is a small quantity anyway. 
% 
%\bibliography{paper_alice}

\end{document}